\documentclass[technote,10pt]{IEEEtran}
\usepackage{cite}
\usepackage{amsmath}
\usepackage{amssymb}
\usepackage{outlines}
\usepackage{array}
\usepackage{float}
\usepackage{kotex}
\usepackage{graphicx}
\usepackage{bm}
\usepackage{caption,subcaption}
\usepackage{algorithm}
\usepackage{algpseudocode}
\usepackage{csquotes}

\begin{document}

\title{Hybrid UAV-enabled Secure Offloading\\ via Deep Reinforcement Learning}

\author{Seonghoon~Yoo, Seongah~Jeong,~\IEEEmembership{Member,~IEEE,} Joonhyuk~Kang,~\IEEEmembership{Member,~IEEE}

\thanks{Seonghoon Yoo is with the Department of Electrical Engineering, Korea Advanced Institute of Science and Technology, Daejeon 34141, South Korea (e-mail: shyoo902@kaist.ac.kr).}
\thanks{Seongah Jeong is with the School of Electronics Engineering, Kyungpook
National University, Daegu 14566, Korea (e-mail: seongah@knu.ac.kr).}
\thanks{Joonhyuk Kang is with the Department of Electrical Engineering, Korea Advanced Institute of Science and Technology, Daejeon 34141, South Korea (e-mail: jhkang@ee.kaist.ac.kr).}
}

{}

\maketitle
\begin{abstract}
Unmanned aerial vehicles (UAVs) have been actively studied as moving cloudlets to provide application offloading opportunities and to enhance the security level of user equipments (UEs). In this correspondence, we propose a hybrid UAV-aided secure offloading system in which a UAV serves as a helper by switching the mode between jamming and relaying to maximize the secrecy sum-rate of UEs. This work aims to optimize (\romannumeral 1) the trajectory of the helper UAV, (\romannumeral 2) the mode selection strategy and (\romannumeral 3) the UEs’ offloading decisions under the constraints of offloading accomplishment and the UAV’s operational limitations. The solution is provided via a deep deterministic policy gradient (DDPG)-based method, whose superior performance is verified via a numerical simulation and compared to those of traditional approaches.
\end{abstract}


\IEEEpeerreviewmaketitle

\section{Introduction}
Recently, unmanned aerial vehicles (UAVs) have begun to play an important role as moving cloudlets for edge computing thanks to their high flexibility and mobility. In particular, UAVs are employed to provide task offloading opportunities beyond 5G and 6G services with high-complexity and low-latency requirements \cite{oma}, \cite{Seongah}. The joint design of offloading resource allocation and the UAV trajectory is proposed in \cite{Seongah} to minimize energy consumption.

With the frequent appearance of the line-of-sight (LoS) paths in the offloading systems via UAV-mounted cloudlets, maintaining privacy and security is challenging. To resolve this issue, physical-layer security technologies have been explored \cite{full_duplex}, \cite{secure_uav_mec}. In \cite{full_duplex}, a full-duplex legitimate UAV acting as an edge server is developed with the optimal design of jamming and user association. The authors in \cite{secure_uav_mec} propose an energy-efficient offloading procedure for a UAV-assisted secure edge computing system with the aim of minimizing the energy consumption of the UAV's data processing. Both \cite{full_duplex} and \cite{secure_uav_mec} provide conventional mathematical solutions, which require adaptive updating according to the time-variant offloading environment, e.g., the channel condition, and therefore encounter the computational complexity issue with an increase in the number of users. To address the complexity of the mathematical approaches, deep reinforcement learning (DRL) has emerged as a promising solution. DRL-based secure transmission in UAV-assisted mobile edge computing is developed in \cite{secure_DRL_offload} to maximize the system utility function. Other authors \cite{smart} propose the optimal design of the legitimate UAV trajectory, the user's transmit power and scheduling for secure communication by adopting a deep deterministic policy gradient (DDPG)-based method, a type of DRL method that can be used to solve continuous control problems. The existing DRL-based methods \cite{secure_DRL_offload}, \cite{smart} for secure offloading focus on the deployment or trajectory design of the legitimate UAV, in the former case of which the operation mode defaults to a single role, such as relaying or jamming.

In this correspondence, we propose a hybrid UAV-aided secure offloading scheme in which a UAV is employed as a helper, switching the mode between jamming and relaying in order to maximize the secrecy sum-rate of user equipments (UEs). The objective of this work is to optimize (i) the trajectory of the helper UAV, (ii) mode selection strategy, and (iii) the UEs’ offloading decisions under the constraints of offloading accomplishment and the UAV’s operational limitations. To this end, we formulate the problem based on a Markov decision process (MDP), whose solution is provided via a DDPG-based method. Via numerical results, the superior performance of the proposed algorithm is verified and compared to those of conventional approaches. 

\section{System Model}
\begin{figure}[t] 
\begin{center}
\includegraphics[width=0.9\columnwidth]{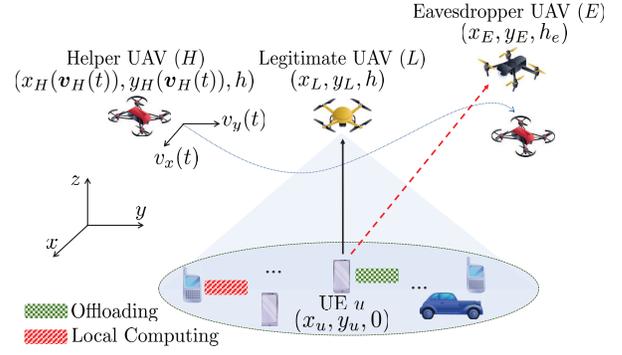}
\caption{Illustration of hybrid UAV-enabled secure offloading.}
\label{System model}
\end{center}
\vspace{-10pt}
\end{figure}

\begin{table*}[t]
\centering
\caption{The achievable rates of a legitimate UAV and eavesdropper UAV according to the mode of the helper UAV}
\vspace{-5pt}
\label{rate}
\scalebox{0.9}{
\begin{tabular}{c|c|c}
\noalign{\smallskip}\noalign{\smallskip}\hline\hline
& Relay mode & Jamming mode \\
\hline
\\
$R^d_u\big(k(t), \boldsymbol{v}_H(t)\big)$  & $\dfrac{1}{2}\min \bigg\{ \log_2 \bigg(1+\dfrac{p_H(k(t)) g_{H,L}(\boldsymbol{v}_H(t))+p_u g_{u,L}(t)}{\sigma^2}\bigg),\log_2 \bigg(1+\dfrac{p_u g_{u,H}(\boldsymbol{v}_H(t))}{\sigma^2}\bigg) \bigg \}$ & $\log_2 \bigg(1+\dfrac{p_ug_{u,L}(t)}{p_H(k(t))g_{H,L}(\boldsymbol{v}_H(t))+\sigma^2} \bigg)$ \\
\hline
\\
$R^e_u\big(k(t),\boldsymbol{v}_H(t)\big)$ & $\dfrac{1}{2}\log_2 \bigg(1+\dfrac{p_H(k(t))g_{H,E}(\boldsymbol{v}_H(t))+p_ug_{u,E}(t)}{\sigma^2} \bigg)$ & $\log_2 \bigg( 1+\dfrac{p_ug_{u,E}(t)}{p_H(k(t))g_{H,E}(\boldsymbol{v}_H(t))+\sigma^2} \bigg)$ \\
\hline
\hline
\end{tabular}}
\vspace{-10pt}
\end{table*}
We consider a hybrid UAV-enabled secure offloading system in which one legitimate UAV is employed as an edge server for ground UEs, while a helper UAV is adopted as a hybrid node to switch roles between relaying and jamming against a single eavesdropper UAV, as shown in Fig. \ref{System model}. For simplicity and tractability, we assume a pair consisting of a legitimate UAV and a helper UAV in a single cell and focus on the uplink scenario. 
The helper UAV in relay mode assists with communication from the UE to the legitimate UAV by forwarding the offloaded data, in which the decode-and-forward (DF) method \cite{ts_frac} is considered. 
In jamming mode, the helper UAV generates artificial noise against the eavesdropper UAV. Here, mode selection at the helper UAV is assumed to be optimized for each time slot (TS). 
According to the helper UAV's role, the legitimate UAV receives the offloaded data from both the helper UAV and UEs in relay mode or from UEs in jamming mode and executes the computation of the received offloaded data. To provide stability during offloading procedure, we assume that the legitimate UAV is hovering with a fixed altitude to serve all UEs within its coverage. 

In the following, we denote the legitimate UAV as $L$, the helper UAV as $H$, and the eavesdropper UAV as $E$. A pair consisting of the legitimate UAV and the helper UAV hovers at an altitude $h$, and the eavesdropper UAV hovers at a higher altitude $h_e$. The $U$ UEs transmit the data to the legitimate UAV for offloading with the orthogonal multiple access. The time horizon $N$ is divided into $T$ TSs, each of which has $\Delta$ seconds, i.e., $N=T\Delta$. For the orthogonal access of multiple UEs, each $\Delta$ is divided equally for the number $U$ of UEs, i.e., $\Delta/U$ seconds of each slot is allocated to each UE. In TS $t$, the mode of the helper UAV is denoted as\\
\vspace{-5pt}
\begin{equation}
    k(t)=\begin{cases}
    1, \ \textrm{if the helper UAV is in relay mode}\\
    0, \ \textrm{if the helper UAV is in jamming mode}.\\
  \end{cases}
\vspace{-5pt}
\end{equation}\\
Also, we define the offloading decision variable $z_{u}(t)$ of the UE $u$ as\\
\begin{equation}
    z_{u}(t)=\begin{cases}
    1, \ \textrm{if the UE $u$ performs offloading}\\
    0, \ \textrm{if the UE $u$ performs local execution}\\
  \end{cases}\hspace{-10pt}, \forall{u}\in \mathcal{U},
\end{equation}
where the set of UEs is denoted as $\mathcal{U} \triangleq \{1,2,...,U\}$. In TS $t$, the helper UAV flies at a constant velocity in terms of the horizontal velocity 
$v_{x}(t)$ and the vertical velocity $v_{y}(t)$, yielding the set of the helper UAV's velocity variables, defined as $\boldsymbol{v}_H(t)=\{v_{x}(t), v_{y}(t)\}$. Accordingly, the horizontal coordinates of the helper UAV can be expressed as $(x_H(\boldsymbol{v}_H(t)),y_H(\boldsymbol{v}_H(t)))$ while satisfying $x_H(\boldsymbol{v}_H(t))=x_H(0)+\sum_{t'=1}^{t}v_{x}(t')\Delta$ and $y_H(\boldsymbol{v}_H(t))=y_H(0)+\sum_{t'=1}^{t}v_{y}(t')\Delta$, both of which are limited by its maximum velocity $v_{\max}$. The legitimate UAV, the eavesdropper UAV and the UE $u$ are assumed to be located on the $xy$-plane at $(x_L,y_L)$, $(x_E,y_E)$ and $(x_{u},y_{u})$, respectively.

By following \cite{rician_12}, Rician fading is adopted for the ground-to-air (G2A) channel, and therefore the channel power gain between UE $u$ and UAV $i$ in TS $t$ can be written as 
\begin{equation}
    \begin{aligned}
        g_{u,i}(\boldsymbol{v}_H(t))=&\dfrac{\beta_0}{(h_{u,i})^2+\big(D_{u,i}\big)^2}\gamma^\textrm{G2A}(t),
    \end{aligned}
\end{equation}
for $\forall u \in \mathcal{U}$ and $i\in \{H,L,E\}$, where $D_{u,i}$ represents the Euclidean horizontal distance between UE $u$ and UAV $i$ on the $xy$-plane as $D_{u,i}\hspace{-2pt}=\hspace{-2pt}\sqrt{\big(x_u-x_i\big)^2+\big(y_u-y_i\big)^2}$, $h_{u,i}$ represents the altitude of UAV $i$ and is defined as $h$ and $h_e$ when $i\hspace{-2pt}\in\hspace{-2pt}\{H,L\}$ and $i\hspace{-2pt}=\hspace{-2pt}E$, respectively, and $\beta_0$ denotes the received power at the reference distance $d_0\hspace{-1pt}=\hspace{-1pt}1$ m of the G2A link. Also, $\gamma^\textrm{G2A}(t)$ is a small scale fading component in the G2A environment with the $K^{\textrm{G2A}}$ factor defined as $\gamma^\textrm{G2A}(t)=\sqrt{{K^{\textrm{G2A}}}/{(K^{\textrm{G2A}}+1)}}\gamma+\sqrt{{1}/{(K^{\textrm{G2A}}+1)}}\Tilde{\gamma}$  \cite{rician_12}, \cite{rician_eq}, where $\gamma$ denotes the deterministic LOS component with $|\gamma|=1$ and $\Tilde{\gamma}$ 
is a circularly symmetric complex Gaussian (CSCG) random variable. Note that in (3), we explicitly express the dependency of the distance on the UAV's velocity when $i=H$ as $D_{u,i}(\boldsymbol{v}_H(t))$. For the air-to-air (A2A) channel gain between the helper UAV and UAV $i \in \{L,E\}$, we define $g_{H,i}(\boldsymbol{v}_H(t))$ as
\begin{equation}
\begin{aligned}
  g_{H,i}(\boldsymbol{v}_H(t))=&\frac{\beta_1}{\big(h_{H,i}\big)^2+\big(D_{H,i}(\boldsymbol{v}_H(t))\big)^2}\gamma^\textrm{A2A}(t),
\end{aligned}
\end{equation}
for $i\in\{L,E\}$, where $D_{H,i}(\boldsymbol{v}_H(t))$ is the horizontal distance between helper UAV $H$ and the legitimate or eavesdropper UAV, $h_{H,i}$ represents the altitude difference, and is defined as $0$ if $i=L$ or as $h-h_e$ if $i=E$, $\beta_1$ denotes the reference channel power gain of the A2A link, and $\gamma^\textrm{A2A}(t)$ is small scale fading component with the $K^{\textrm{A2A}}$ factor \cite{rician_20}. Since the UE $u$ can offload the data to the legitimate UAV or can be supported by the helper UAV within their coverage area in TS $t$, we have
\begin{equation}
z_{u}(t)\big(k(t)D_{u,i}+(1-k(t))D_{u,L}\big)\leq D_{\max},\vspace{-3pt}
\end{equation}
for $\forall u \in \mathcal{U}$ and $i \in \{H,L\}$, where $D_{\max}$ denotes the radius of the coverage area for both the legitimate and helper UAVs.
\subsection{Communication Model}
In this section, we provide the communication model required for the secure offloading procedure between the legitimate UAV and UEs. For the relay operation of the helper UAV, we adopt a the time division manner due to the half-duplex limitation \cite{ts_frac}. In particular, the time fraction $\Delta/U$ allocated to each UE is divided into two parts, the first of which is used for each UE to transmit the data to both the legitimate and helper UAV, while the remainder is adopted for the helper UAV to relay the received data to the legitimate UAV. In jamming mode, the entire $\Delta/U$ is consumed for transmission from each UE to the legitimate UAV while the helper UAV generates the jamming signal. 
Depending on the helper UAV's operation mode, the achievable data rates $R^d_u\big(k(t), \boldsymbol{v}_H(t)\big)$ and $R^e_u\big(k(t), \boldsymbol{v}_H(t)\big)$ at the legitimate UAV and eavesdropper UAV are calculated as in Table \ref{rate}. In Table \ref{rate}, $p_u$ is the transmit power of UE $u$, $\sigma ^2$ is the noise power, and $p_H(k(t))$ is the transmit power of the helper UAV, where $p_H(k(t)) =p_R$ in relay mode, otherwise $p_H(k(t)) =p_J$. For a further performance gain, the optimal power allocation for a different TS can be considered, which is left as our future work. Note that the achievable data rate at the legitimate UAV in relay mode is expressed as the minimum data rate obtained in two time fractions of the DF protocol, while the eavesdropper UAV can overhear the data via both the UE-legitimate UAV link and the helper UAV-legitimate UAV link. In jamming mode, the artificial interference at the eavesdropper UAV caused by the friendly jamming of the helper UAV is factored into the data rate. Consequently, the secrecy sum-rate of the wiretap channel is written as
\vspace{-4pt}
\begin{equation}
\begin{aligned}
    &C\big(k(t), \boldsymbol{z}(t), \boldsymbol{v}_H(t)\big)=\vspace{-3pt}\\&\sum\limits_{u\in\mathcal{U}}z_u(t)\big[R^d_{u}\big(k(t), \boldsymbol{v}_H(t)\big)-R^e_{u}\big(k(t), \boldsymbol{v}_H(t)\big)\big]^+,
\end{aligned}
\vspace{-5pt}
\end{equation}
where $[x]^+\triangleq\max(x,0)$, $\boldsymbol{z}(t) = \{z_u(t)\}_{u \in \mathcal{U}}$. 
\vspace{-3pt}

\subsection{Computing Model}
We define the computational task of UE $u$ in TS $t$ as $\{S_{u}(t),F_{u}(t) \}$, where $S_{u}(t)$ denotes the data size of the task, and $F_{u}(t)$ denotes the number of CPU cycles for computing one bit.
When the UE does the local execution, the task is computed within $\Delta$, and hence the CPU frequency $f_u(t)$ of the UE $u$ is determined as $f_u(t) = S_u(t)F_u(t)/\Delta$. At the legitimate UAV, the total data received at the previous TS $t-1$ is assumed to be computed in TS $t$, and the CPU frequency of the legitimate UAV, $f_L(\boldsymbol{z}(t))$, is calculated as $f_L(\boldsymbol{z}(t))={\sum\limits_{u \in \mathcal{U}}z_u(t-1)S_u(t-1)F_u(t-1)}/{\Delta}$.

\subsection{Energy Model}
Here, since all network components have limited battery capabilities, their energy consumption needs to be addressed in the system design phase. 
The computation energy $E^C_{i}(\boldsymbol{z}(t))$ required for execution at $i \in \{1,...,U,L\}$ is given by \cite{full_duplex}
\begin{equation}
\vspace{-3pt}
    E^C_{i}(\boldsymbol{z}(t))=\kappa(f_{i})^3\Delta,\\
\end{equation}
where $\kappa$ denotes the power consumption coefficient, and the CPU frequency $f_{i}$ is substituted with $f_u(t)$ and $f_L(\boldsymbol{z}(t))$ when $i \in \mathcal{U}$ and $i=L$, respectively. The energy consumption at the helper UAV results from the signal transmission 
and the flying operation. The transmission energy consumption is derived as $E^{Tr}_H(k(t),\boldsymbol{z}(t))\hspace{-3pt}=\hspace{-2pt}k(t)\sum_{u \in \mathcal{U}}z_u(t)p_R{\Delta}/{(2U)}+(1-k(t))p_J\Delta$, while the flying energy consumption is given via $E^F_H(\boldsymbol{v}_H(t))=0.5M\Delta\big((v_x(t))^2+(v_y(t))^2\big)$ \cite{Seongah}, where $M$ is the mass of the UAV, including its payload.

\section{Proposed DDPG-based method}

This work aims to maximize the secrecy sum-rate by jointly optimizing the helper UAV's mode $k(t)$, the UE's offloading choice $\boldsymbol{z}(t)$ and the helper UAV's velocity $\boldsymbol{v}_H(t)$ for all $t$. To this end, we formulate the optimization problem as follows: 
\vspace{-5pt}
\begin{subequations}\label{eq}
\begin{eqnarray}
&&\hspace{-1.2cm}\underset{k(t), \boldsymbol{z}(t), \boldsymbol{v}_H(t)}{\text{max}} C(k(t), \boldsymbol{z}(t), \boldsymbol{v}_H(t))\\[2pt]
&&\hspace{-1cm}\text{s.t.}\hspace{+0.6cm} k(t)=\{0,1\}, \,\,z_{u}(t)=\{0,1\}, \quad \forall u \in \mathcal{U},\label{eqb}\\[2pt]
&& -l_{\max}/2 \leq x_H(\boldsymbol{v}_H(t)), y_H(\boldsymbol{v}_H(t))\leq l_{\max}/2, \label{eq{d,e}}\\[2pt]
&& \begin{aligned}
    z_{u}(&t)\big(k(t)D_{u,i}+(1-k(t))D_{u,L}\big)\leq D_{\max}, \\ &\, i\hspace{-1pt}\in\hspace{-1pt}\{H,L\}, \, \forall u \in \mathcal{U}, \label{eqf}
\end{aligned}\\[2pt]
&& \begin{aligned}
    z_{u}(&t)p_u\big(k(t)\frac{\Delta}{2U}\hspace{-2pt}+\hspace{-2pt}(1-k(t))\frac{\Delta}{U}\big)\hspace{-2pt}\\ &+\hspace{-2pt}(1-z_u(t))E^C_{u}(t) \leq E_u, \, \forall u \in \mathcal{U}, \label{eqg}
\end{aligned} \\[2pt]
&& E^C_{L}(\boldsymbol{z}_t) \leq E_L, \label{eqh}\\[2pt]
&& E^F_H(\boldsymbol{v}_H(t))\hspace{-2pt}+\hspace{-2pt}E^{Tr}_H(k(t),\boldsymbol{z}(t)) \leq E_H, \label{eqi}
\end{eqnarray}
\end{subequations}
where (8b) is a binary variable constraint pertaining to the helper UAV's mode and offloading decision, (8c) ensures that the helper UAV travels within a $l_{\max}$-side-length square, (8d) restricts the legitimate and helper UAVs to hover within their coverage area, and (8e)-(8g) represent the energy constraints of UEs, the legitimate UAV and the helper UAV, respectively.

To solve problem (8), we employ the DRL framework to find the optimal policy for $\{k(t), \boldsymbol{z}(t), \boldsymbol{v}_H(t)\}_{\forall t}$ in every TS. Since a real-time mathematical approach for UAV trajectory design has complexity issues, we adopt the DDPG method among DRL-based approaches, which is appropriate for controlling a continuous action space \cite{ddpg}. In the MDP, the agent has state $s_t$ in the environment during discrete TS, and it takes action $a_t$ every TS. As the agent proceeds with various interactions in the environment, the agent obtains a reward $r_t$ and next state $s_{t+1}$. The policy ($\pi$) is designed to maximize the accumulated reward $R_t=\sum_{i=t}^{T}\gamma^{(i-t)}r_i$, where $\gamma \in [0,1]$ is the discount factor. The critic network learns the action-value function $Q(s_t,a_t)=\mathbb{E}_{a_{i>t} \sim \pi}[R_t|s_t,a_t]$ using Bellman's equation in Q-learning and proceeds to minimize the loss function $L(\cdot)$, which is defined as
\vspace{-3pt}
\begin{equation}
    L(\theta^Q)=\mathbb{E}\big[\big(Q\big(s_t,a_t|\theta^Q\big)-y_t\big)^2\big],
\vspace{-2pt}\end{equation}\vspace{-1pt}where $\theta^Q$ is the weight of the critic network and $y_t=r_t+{\gamma}Q'\big(s_{t+1},\mu'(s_{t+1}|\theta^{\mu'})|\theta^{Q'}\big)$. The actor network updates with the policy gradient method to maximize the expected reward $J=\mathbb{E}_{a_i \sim \pi}[R_1]$ and uses a policy function approximator, which follows 
\vspace{-3pt}
\begin{equation}
    \nabla_{\theta^{\mu}}J \approx \mathbb{E}\big[\nabla_a Q(s,a|\theta^Q)|_{s=s_t, a=\mu(s_t)}\nabla_{\theta^{\mu}}\mu(s|\theta^{\mu})|_{s_t}\big],
\end{equation}
where $\theta^\mu$ is the weight factor of the actor network. Additionally, the DDPG algorithm improves the update stability by using the target networks $\theta^{Q'}$ and $\theta^{\mu'}$, which are identical to those of the critic network and the actor network, and these target networks are updated by a soft update method.

To optimize the helper UAV's trajectory modeled as an MDP, we define the state, action and reward function in TS $t$ as follows:

\textbf{State: }Let $\mathcal{S}$ denote the system state space as $\mathcal{S}=\{s_t|s_t=\{x_H(\boldsymbol{v}_H(t)), y_H(\boldsymbol{v}_H(t)),k(t),  \{D_{H,i}(\boldsymbol{v}_H(t))\}_{i\in \mathcal{U} \cup \{L,E\}}\}, \\t\in\{1,2,...,T\}\}$, whose components are the coordinates and the mode of the helper UAV, and the horizontal distance between the helper UAV and other nodes, respectively. 

\textbf{Action: }Let $\mathcal{A}$ denote the system action space as $\mathcal{A}=\{a_t|a_t=\{v_{x}(t),  v_{y}(t)\}, t\in \{1,2,...,T\}\}$,
whose components are the horizontal and vertical velocity of the helper UAV.

\textbf{Reward: }We define $r_t = C(k(t), \boldsymbol{z}(t), \boldsymbol{v}_H(t))\hspace{-1pt}-\hspace{-1pt}r_{om}$ as a reward function focusing on maximization of the secrecy sum-rate, where $r_{om}$ is the penalty value for cases in which the helper UAV goes off of the given map. Note that the helper UAV returns to the previous location if it goes off of the map.

With the optimized helper UAV's trajectory, we develop a relaxation method to optimize the offloading decision variable $\boldsymbol{z}(t)$ and the helper UAV's operation mode $k(t)$. The offloading decision variable $\boldsymbol{z}(t)$ is designed as \begin{equation}
    z_u(t)=\begin{cases}
    \!1,& \begin{aligned} \hspace{-5pt}\text{if} \hspace{3pt} R_u^d\big(k(t), \boldsymbol{v}_H(t)\big)\hspace{-2pt}-\hspace{-2pt}R_u^e\big(k(t),\boldsymbol{v}_H(t)\big)\hspace{-2pt}>\hspace{-2pt}\varepsilon, \\
       & \hspace{-170pt}\forall{u} \in \mathcal{U}\text{ and (5) are satisfied}
    \end{aligned}            \\
    0, &  \hspace{-5pt}\text{otherwise,} 
  \end{cases}\hspace{80pt}
\end{equation} so that a secrecy sum-rate greater than the minimum limit $\varepsilon$ is guaranteed, and (5) is satisfied after action $a_t$ is performed. 
According to the offloading decision, the helper UAV's mode for providing a higher secrecy sum-rate is selected by $k(t)={\arg\!\max}_{i\in\{0,1\}}C_i$, and the reward $r_t$ is defined as $r_t={\max}_{i\in\{0,1\}}C_i-r_{om}$.

Based on the entire process mentioned above, we propose the DDPG-based method, as given in Algorithm 1. In order to increase the convergence speed of {Algorithm 1}, the initial weights are set experimentally based on the previous steps, where a higher reward is achieved. \begin{algorithm}[t]
	\caption{DDPG-based method for a hybrid UAV-enabled secure offloading system}
	\hspace*{\algorithmicindent} \textbf{Input:} Structures of the actor, critic and target network.
	\begin{algorithmic}[1]
    \State \textbf{Initialize: }Actor $\mu$, critic $Q$ and target network $\mu'$, $Q'$  with weights $\theta^\mu$, $\theta^Q$ and $\theta^{\mu'}\hspace{-4pt}\leftarrow\hspace{-1pt}\theta ^{\mu}$, $\theta^{Q'}\hspace{-4pt}\leftarrow\hspace{-1pt}\theta^Q$ and replay buffer $\mathcal{B}$;
	\For {TS in $T$}
	    \State Set $\{{z}_u(t)=0\}_{\forall{u} \in \mathcal{U}}$, and $C_0=C_1=0$;
	    \State Calculate $s_t$ as in \enquote{State} step of MDP;
	    \State Execute action $a_t=\mu(s_t|\theta^\mu)+\mathcal{N}$; 
	    \For {$i \in \{0,1\}$}
	        \State $k(t) = i$;
	        \State Obtain the offloading decision $\boldsymbol{z}(t)$ by (11);
	        \State Calculate $C_i=C(k(t),\boldsymbol{z}(t), \boldsymbol{v}_H(t))$
	    \EndFor
	    \State Obtain reward $r_t\hspace{-2pt}=\hspace{-3pt}\max\limits_{i\in\{0,1\}}C_i-r_{om}$ and next state $s_{t+1}$;
	    \State Store transition $\big(s_t, a_t, r_t, s_{t+1}\big)$ in $\mathcal{B}$;
	    \State Sample a random mini-batch of $K$ transitions $\big(s_i, a_i$, 
	    
	    $r_i, s_{i+1}\big)$ from $\mathcal{B}$;
	    \State Update critic, actor network according to (9), (10);
	    \State Update target networks:
	    
	    \begin{center}
	    $\theta^Q \leftarrow \tau \theta^Q+(1-\tau)\theta^{Q'}$, $\theta^{\mu'} \leftarrow \tau \theta^{\mu}+(1-\tau)\theta^{\mu'}$;
	    \end{center}
	\EndFor
	\end{algorithmic} 
	\hspace*{\algorithmicindent} \textbf{Output:} Actor network $\mu(s_t|\theta^\mu)$.
\end{algorithm} For the helper UAV, action $a_t$ is generated by the actor network $\mu$, and a noise process $\mathcal{N}$ is added for exploration. 
Then, we obtain the reward $r_t$ and next state $s_{t+1}$, while the helper UAV stores the transition into its finite-sized buffer $\mathcal{B}$. From Line 14 to 15, networks are updated by pulling $K$ samples from the buffer.

\section{Simulation Results}
In this section, we present the numerical results to evaluate the performance of the proposed algorithm compared to the reference methods. For simulations, we consider the parameter settings shown in Table \ref{parameter} by following \cite{full_duplex}, \cite{RL_Hanzo}. For the energy budget of each node, we set $E_u=0.025$J, $E_L=24$J, and $E_H$ is set to $3.9$KJ \cite{energy_budget}. In addition, we set 1000 episodes in the training stage. 
The capacity of the replay buffer is 8000, and the mini-batch size is 70. The noise process $\mathcal{N}$ follows a normal distribution with a zero mean and variance of 0.6. Noise decays at a rate of 0.999.
The actor and critic networks 
have three fully-connected hidden layers with [300,100,100] neurons, and are trained at a learning rate of $10^{-4}$. The activation function is used as tanh function, and the network is updated using the AdamOptimizer. For references, the following benchmark methods are considered:
\vspace{-2pt}
\begin{outline}
    \1 Relay mode with linear trajectory (Re-LT): Scheme with a linear trajectory to reach the midpoint between the legitimate UAV and UEs at TS $T$ based on proposed offloading decision method in relay mode of the helper UAV.
    \1 Jamming mode with linear trajectory (Ja-LT): Scheme with a linear trajectory to reach the point, where the eavesdropper UAV exists at TS $T$ based on proposed offloading decision method in jamming mode of the helper UAV.
    \1 Relay mode with optimal trajectory (Re-OT): Scheme with an optimal trajectory based on proposed offloading decision method in relay mode of the helper UAV.
    \1 Jamming mode with optimal trajectory (Ja-OT): Scheme with an optimal trajectory based on proposed offloading decision method in jamming mode of the helper UAV.
\end{outline}
\begin{table}[t]
    \scriptsize
    \vspace{-2pt}
    \caption{Simulation parameters}
    \label{parameter}
    \centering
    \vspace{-4pt}
    \begin{tabular}{cccccc} 
        \\[-1.8ex]\hline 
        \hline \\[-1.8ex] 
        \multicolumn{1}{c}{Parameter} & \multicolumn{1}{c}{Value} & \multicolumn{1}{c}{Parameter} & \multicolumn{1}{c}{Value}\\
        \hline \\[-1.8ex] 
        {$U$} & $10$  & {$S_{u}$} &  $[20,30]\;$KB \\
        {$T$} &  $10$, $20$ &{$F_{u}$} &  $[1000,1200]\;$cycles/bit   \\
        {$l_{\max}$} & $200\;$m   &{$p_{u}$} &  $0.1\;$W\\
        {$\Delta$} & $1\;$s  &{$p_{J}$} &  $0.08\;$W\\
        {$h$} &  $80\;$m  &{$p_{R}$} &  $0.012\;$W\\
        {$h_e$} &  $120\;$m   &{$\sigma^2$} &  $-100\;$dBm\\
        {$\varepsilon$} &  $0.1\;$bps/Hz    &{$\beta_0$} &  $10^{-5}$    \\ 
        {$D_{\max}$} &  $45\;$m &{$\beta_1$} &  $10^{-4}$    \\ 
        {$v_{\max}$} &  $20\;$m/s &{$r_{om}$} &  $0.2$    \\ 
        {$\kappa$} &  $10^{-27}$  &  {$\tau$} &  $0.005$     \\ 
        {$K^{\textrm{G2A}}$} &  $12\;$dB \cite{rician_12} & {$\gamma$} &  $0.95$  \\ 
        {$K^{\textrm{A2A}}$} &  $20\;$dB \cite{rician_20} & {$M$} & $9.65\;$kg\\
        \\[-1.8ex]\hline 
        \hline \\[-1.8ex] 
    \end{tabular}
    \vspace{-10pt}
\end{table} 
\begin{figure}[t] 
\vspace{-5pt}
\begin{center}
\includegraphics[width=0.77\columnwidth]{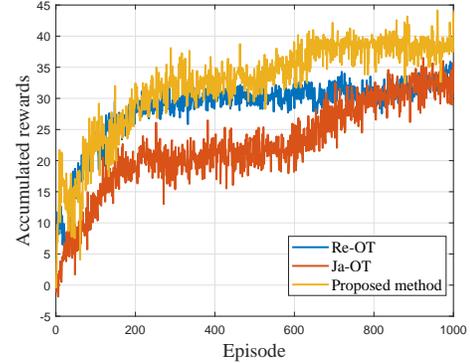}
\caption{Accumulated reward versus episodes.}
\label{Accumulated reward vesus training episodes}
\end{center}
\vspace{-12pt}
\end{figure}

\begin{figure}[t]
\vspace{-15pt}
\centering
\begin{center}
    \begin{subfigure}{0.5\textwidth}
    \centering
    \includegraphics[width=0.77\textwidth]{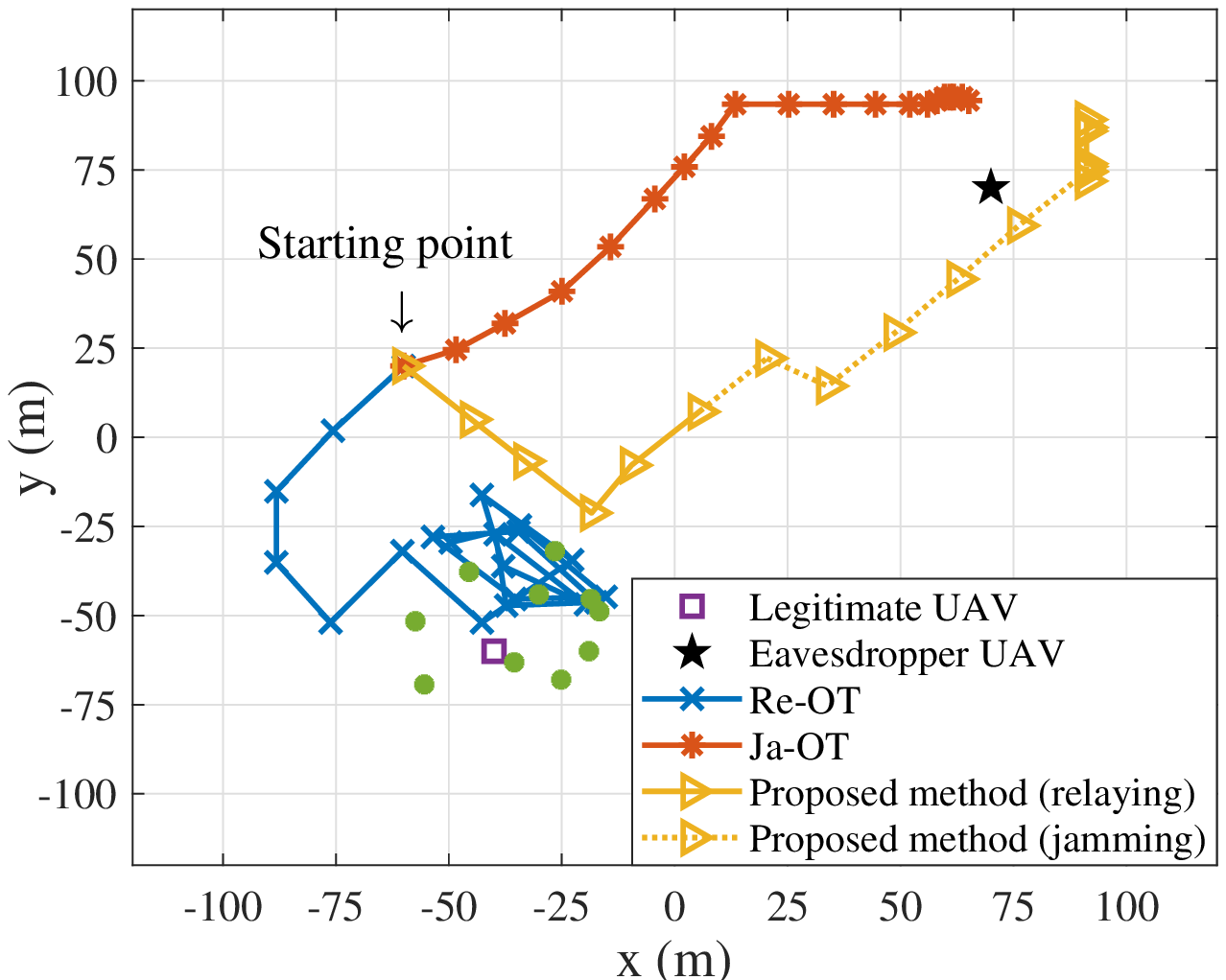}
    \caption{}
\end{subfigure}
\begin{subfigure}{0.5\textwidth}
    \centering
    \includegraphics[width=0.77\textwidth]{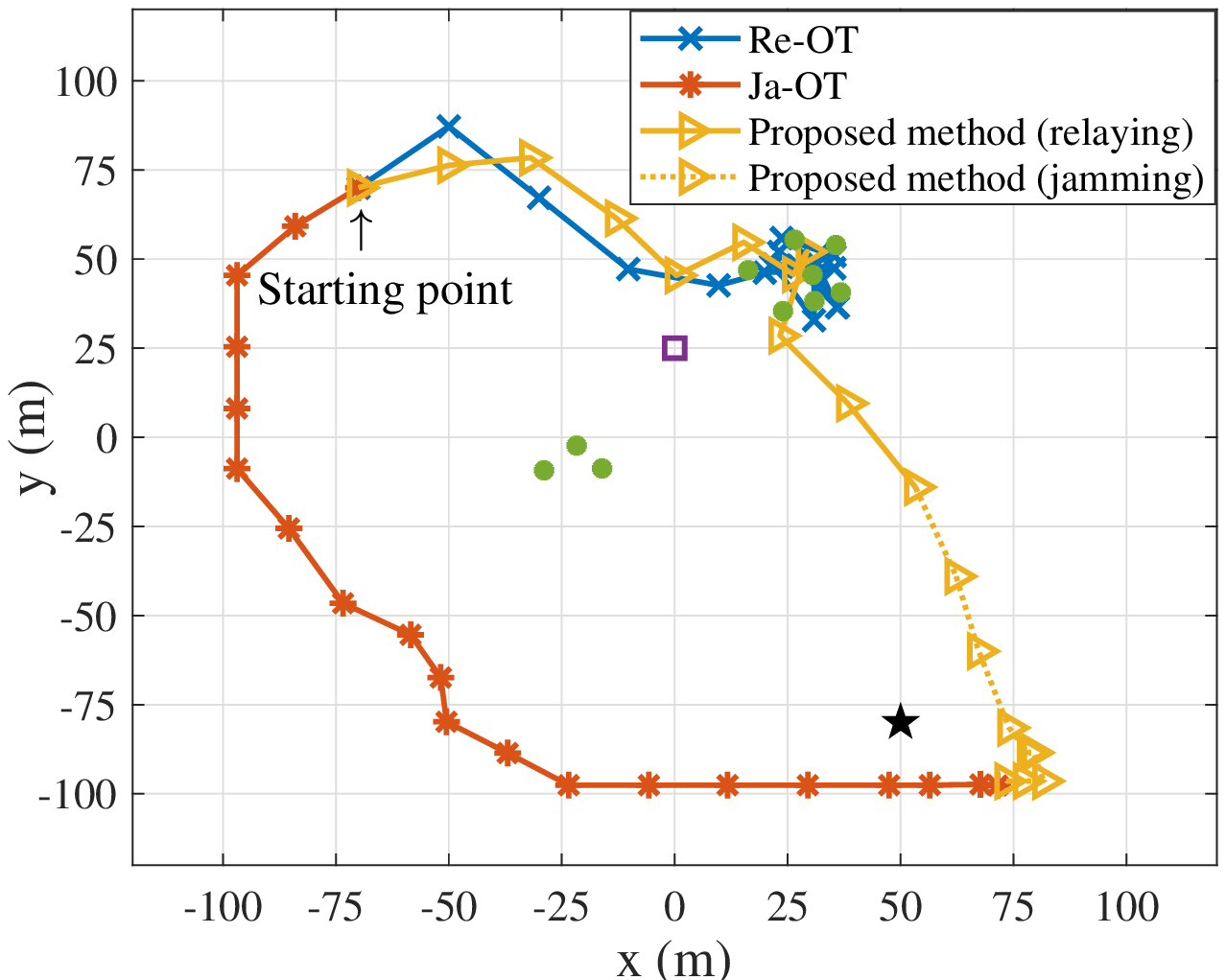}
    \caption{}
\end{subfigure}
\caption{The trajectory of the helper UAV according to the mode during 20 TS. (a) UEs are randomly distributed around the legitimate UAV. (b) Two UE clusters are spatially separated around the legitimate UAV.}
\label{trajectory}
\end{center}
\vspace{-14pt}
\end{figure}

Fig. \ref{Accumulated reward vesus training episodes} shows the accumulated reward of the proposed algorithm as a function of training episodes. It is observed that the proposed method converges after 600 episodes. In addition, the proposed method achieved a higher accumulated reward than the Re-OT and Ja-OT schemes by further optimizing the offloading decision and the operation mode of the helper UAV. Fig. \ref{trajectory} shows the optimal trajectory obtained by the proposed method. 
In Fig. \ref{trajectory}(a), we consider the case in which the UEs are randomly distributed around the legitimate UAV. It is observed that the optimized helper UAV tends to move around the UEs in Re-OT to increase the relay performance, while it moves towards the eavesdropper UAV in Ja-OT to maximize the jamming effect. In the proposed method, the helper UAV initially operates in relay mode (with yellow-solid line) and thus moves toward the UE cluster, similar to Re-OT. From 5 TS (5s), the helper UAV switches to jamming mode (with the yellow-dashed line), and moves toward the eavesdropper UAV, as in Ja-OT. In the case of \ref{trajectory}(b), we consider two spatially separated UE groups around the legitimate UAV. Compared to \ref{trajectory}(a), in Re-OT, the helper UAV moves toward the large-scale cluster with 7 UEs, which can provide a higher secrecy sum-rate. In Ja-OT, the helper UAV goes to the eavesdropper UAV while maintaining its distance from the legitimate UAV. In the proposed method, both movement tendencies in Re-OT and Ja-OT are shown according to the corresponding mode change. 

\begin{figure}[t] 
\vspace{-15pt}
\begin{center}
\includegraphics[width=0.77\columnwidth]{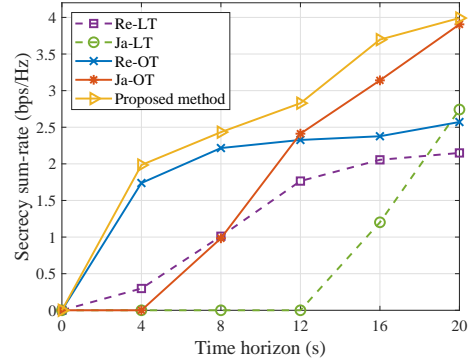}
\caption{The secrecy sum-rate of each schemes versus the time horizon.}
\label{Sumrate}
\end{center}
\vspace{-12pt}
\end{figure}

In Fig. \ref{Sumrate}, the secrecy sum-rate of the proposed method is shown as a function of the different time horizon $N$ in a setting identical to that in Fig. \ref{trajectory}(b). It can be seen that for all schemes, the secrecy sum-rate increases as the mission time increases. Note that the proposed method achieves the best secrecy sum-rate via joint optimization. Moreover, it is obvious that the DDPG-based trajectory design of the helper UAV provides additional secrecy improvements in comparison with Re-OT and Re-LT, or Ja-OT and Ja-LT.
\\

\section{Conclusions}
In this correspondence, we have proposed a hybrid UAV-enabled secure offloading algorithm to maximize the secrecy sum-rate of ground users, where a hybrid helper UAV is adopted to switch roles between relaying and jamming. We jointly optimize the helper UAV's mode selection and trajectory as well as users' offloading decisions based on the the DDPG method. Via simulations, the superior performance of the proposed method is verified compared to those of conventional methods. A scenario with multiple helpers and eavesdroppers can also be studied with the non-orthogonal multiple access in future work.

\ifCLASSOPTIONcaptionsoff
  \newpage
\fi

\bibliographystyle{IEEEtran}
\bibliography{ref}


\end{document}